\documentstyle[multicol,aps,psfig]{revtex}
\renewcommand{\narrowtext}{\begin{multicols}{2}
\global\columnwidth20.5pc\noindent}
\renewcommand{\widetext}{\end{multicols}
\global\columnwidth42.5pc}
\begin{document}
\draft
\preprint{28 June 2002}
\title{Quantum and thermal phase transitions in the Bechgaard salts
       and their analogs}
\author{Hiromitsu Hori$^{\rm a,b}$ and Shoji Yamamoto$^{\rm b}$}
\address
{$^{\rm a}$Department of Physics,
           Okayama University, Okayama 700-8530, Japan}
\address
{$^{\rm b}$Division of Physics,
           Hokkaido University, Sapporo 060-0810, Japan}

\date{Received 28 June 2002}
\maketitle
\begin{abstract}
In order to investigate quantum and thermal phase transitions in the
Bechgaard salts and their sulfur analogs, we perform
finite-temperature Hartree-Fock calculations in one dimension with
particular emphasis on the interplay between charge ordering and
lattice instability.
The coexisting charge- and spin-density-wave state as well as its
precursor fluctuations in $\mbox{(TMTSF)}_2\mbox{PF}_6$ and the
lattice tetramerization in $\mbox{(TMTTF)}_2\mbox{ReO}_4$ are
well interpreted.
\end{abstract}
\pacs{{\it Keywords}:
      Equilibrium thermodynamics and statistical mechanics,
      Order-disorder phase transitions,
      Organic conductors based on radical cation and/or anion salts}
\narrowtext

   The Bechgaard salts $\mbox{(TMTSF)}_2X$ and their sulfur analogs
$\mbox{(TMTTF)}_2X$, where $X=\mbox{PF}_6, \mbox{ReO}_4, \mbox{Br},
\mbox{etc.}$, exhibit various types of charge and/or spin ordering
\cite{P1501} and stimulate our interest in quasi-one-dimensional
correlated electron systems with a three-quarter-filled
(quarter-filled in terms of holes) $\pi$ band.
A variety of mechanisms have been proposed in an attempt to interpret
the variety of instabilities, featuring dimensionality
\cite{M1522,S805,M13400,R16243,R241102}, which can be tuned by
pressure as well as chemically, intrachain dimerization
\cite{R241102,S1249,S235107,K3356} due to the anion columns, and
competing Coulomb interactions
\cite{S1249,S235107,K1098,T125123}$-$on-site, nearest-neighbor, and
even next-nearest-neighbor repulsions.
Recently, introducing another interesting viewpoint$-$the interplay
between charge ordering and lattice instability, several authors
\cite{M1522,R16243,R241102} have revealed further exotic density-wave
states.
In particular, Riera and Poilblanc \cite{R16243,R241102} found
charge-density-wave (CDW)-bond-order-wave (BOW) mixed phases of new
type, which consist of the superposition of lattice dimerization and
tetramerization, and pointed out their relevance to the spin-Peierls
transition \cite{C1698} recently observed in TMTTF salts.

   Thus motivated, we explore thermal, as well as quantum, phase
transitions in the Bechgaard-Fabre salts taking account of long-range
Coulomb and electron-lattice interactions.
We consider a one-dimensional model at quarter filling,
\begin{equation}
   {\cal H}
   ={\cal H}_T+{\cal H}_U+{\cal H}_V+{\cal H}_{V'}+{\cal H}_K\,,
   \label{E:H}
\end{equation}
where
\begin{eqnarray}
   &&
   {\cal H}_T
   =\sum_{j,\sigma}
    \left(
     t_j c_{j,\sigma}^\dagger c_{j+1,\sigma}+{\rm H.c.}
    \right)\,,
   \nonumber \\
   &&
   {\cal H}_U=U\sum_j n_{j,\uparrow}n_{j,\downarrow}\,,
   \nonumber \\
   &&
   {\cal H}_V
   =\sum_j V_j n_j n_{j+1}\,,\ \ 
   {\cal H}_{V'}
   =V'\sum_j n_j n_{j+2}\,,
   \nonumber \\
   &&
   {\cal H}_K
   =\frac{1}{2}
    \sum_j K_j(u_{j+1}-u_j)^2\,,
\end{eqnarray}
\vspace*{-2mm}
\begin{figure}
\centerline
{\mbox{\psfig{figure=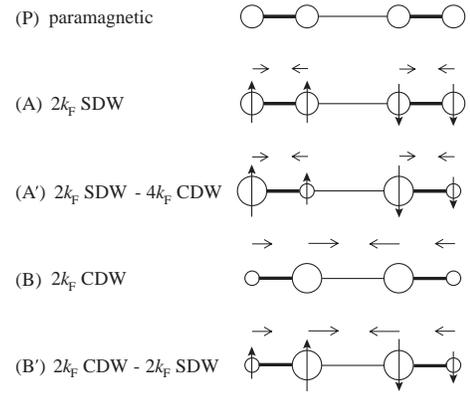,width=60mm,angle=0}}}
\vspace*{3mm}
\caption{Schematic of various spin-charge-lattice instabilities,
         where the various circles, segments, and vertical arrows
         qualitatively describe the variation of charge densities,
         bond orders, and spin densities, respectively, while the
         horizontal arrows denote lattice distortion from
         equilibrium.}
\end{figure}
\noindent
with $c_{j,\sigma}^\dagger$ creating a hole of spin $\sigma$ on the
$j$th molecular site, $u_j$ being the displacement of the $j$th
molecular site from equilibrium,
$n_{j,\sigma}=c_{j,\sigma}^\dagger c_{j,\sigma}$, and
$n_j=n_{j,\uparrow}+n_{j,\downarrow}$.
Considering the intrinsic dimerization of the system due to the
anion columns, we introduce alternation to transfer integrals,
nearest-neighbor Coulomb interactions, and spring constants as
$t_j=-t_\lambda+\alpha_\lambda(u_{j+1}-u_j)$, $V_j=V_\lambda$,
and $K_j=K_\lambda$, where $\lambda$ is set to ``a" for an odd $j$
and to ``b" for an even $j$.
We calculate the free energy within the Hartree-Fock approximation,
which should be distinguished from the mean-field treatment of the
Hartree level \cite{S805,S1249,K3356,K1098} and is crucially superior
to it with intersite Coulomb interactions included.
The lattice distortion is adiabatically treated and can therefore be
described in terms of the electron density matrices.
We always set the smaller ones of the intrinsic transfer integrals
and the spring constants, $t_{\rm b}$ and $K_{\rm b}$, both equal to
unity.
The Hamiltonian (\ref{E:H}) possesses various density-wave solutions,
some of which are schematically shown in Fig. \ref{F:DW}, where (A),
(A$'$), and (B$'$) are accompanied by spin alignment and their
nonlocal stabilization into a spin density wave (SDW) assumes weak
interchain interaction.
\begin{figure}
\centerline
{\mbox{\psfig{figure=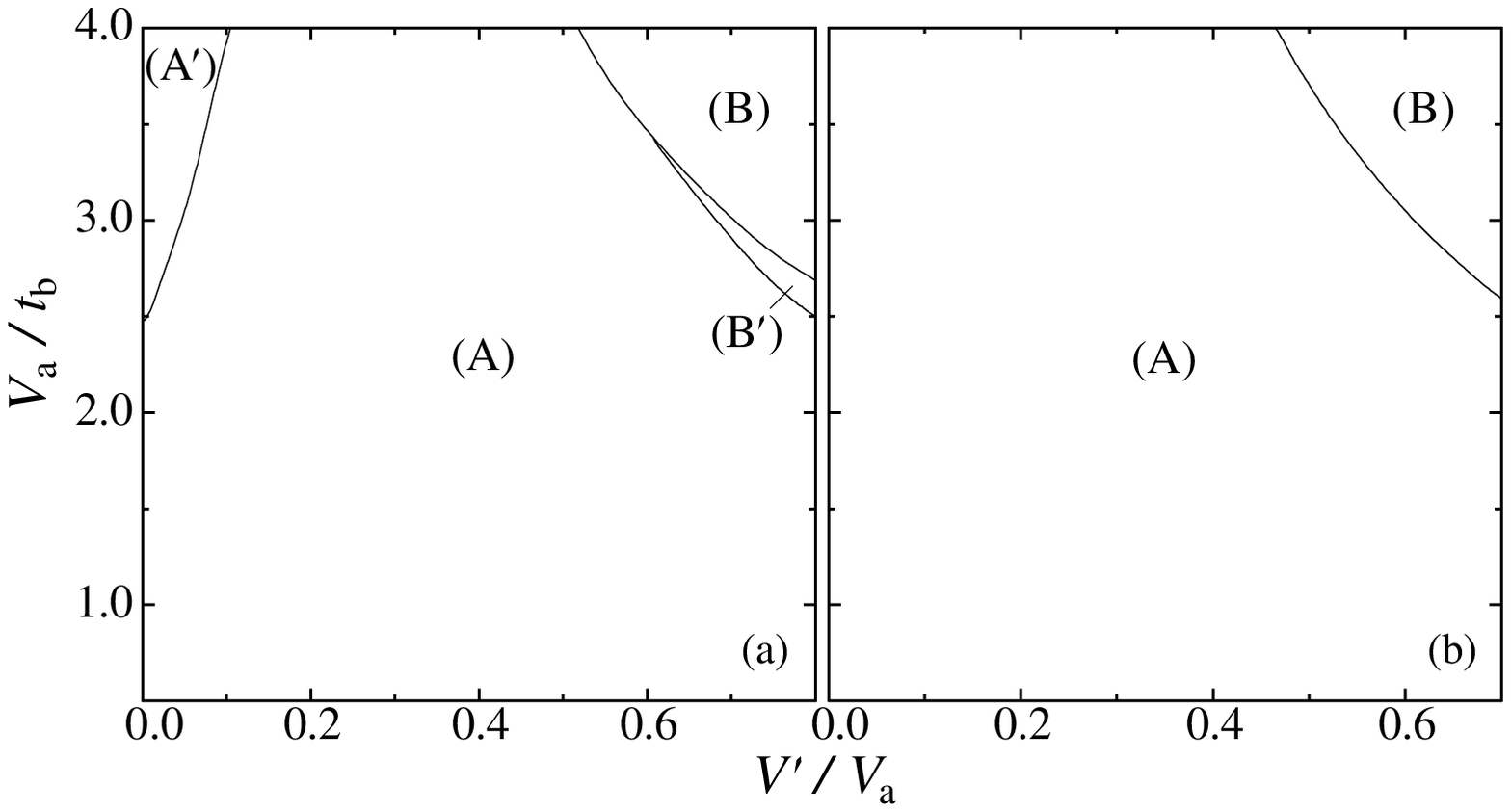,width=80mm,angle=0}}}
\vspace*{1mm}
\caption{Ground-state phase diagrams:
         (a) $t_{\rm a}/t_{\rm b}=1.1$,
             $U/t_{\rm b}=4.0$, $V_{\rm b}/V_{\rm a}=0.8$,
             $\alpha_{\rm a}/(t_{\rm b}K_{\rm a})^{1/2}=0.1$,
             $\alpha_{\rm b}/(t_{\rm b}K_{\rm b})^{1/2}=0.11$,
             $K_{\rm a}/K_{\rm b}=1.1$;
         (b) $t_{\rm a}/t_{\rm b}=1.35$,
             $U/t_{\rm b}=4.0$, $V_{\rm b}/V_{\rm a}=0.7$,
             $\alpha_{\rm a}/(t_{\rm b}K_{\rm a})^{1/2}=0.5$,
             $\alpha_{\rm b}/(t_{\rm b}K_{\rm b})^{1/2}=0.6$,
             $K_{\rm a}/K_{\rm b}=1.2$.}
\label{F:GSPhD}
\vspace*{6mm}
\centerline
{\mbox{\psfig{figure=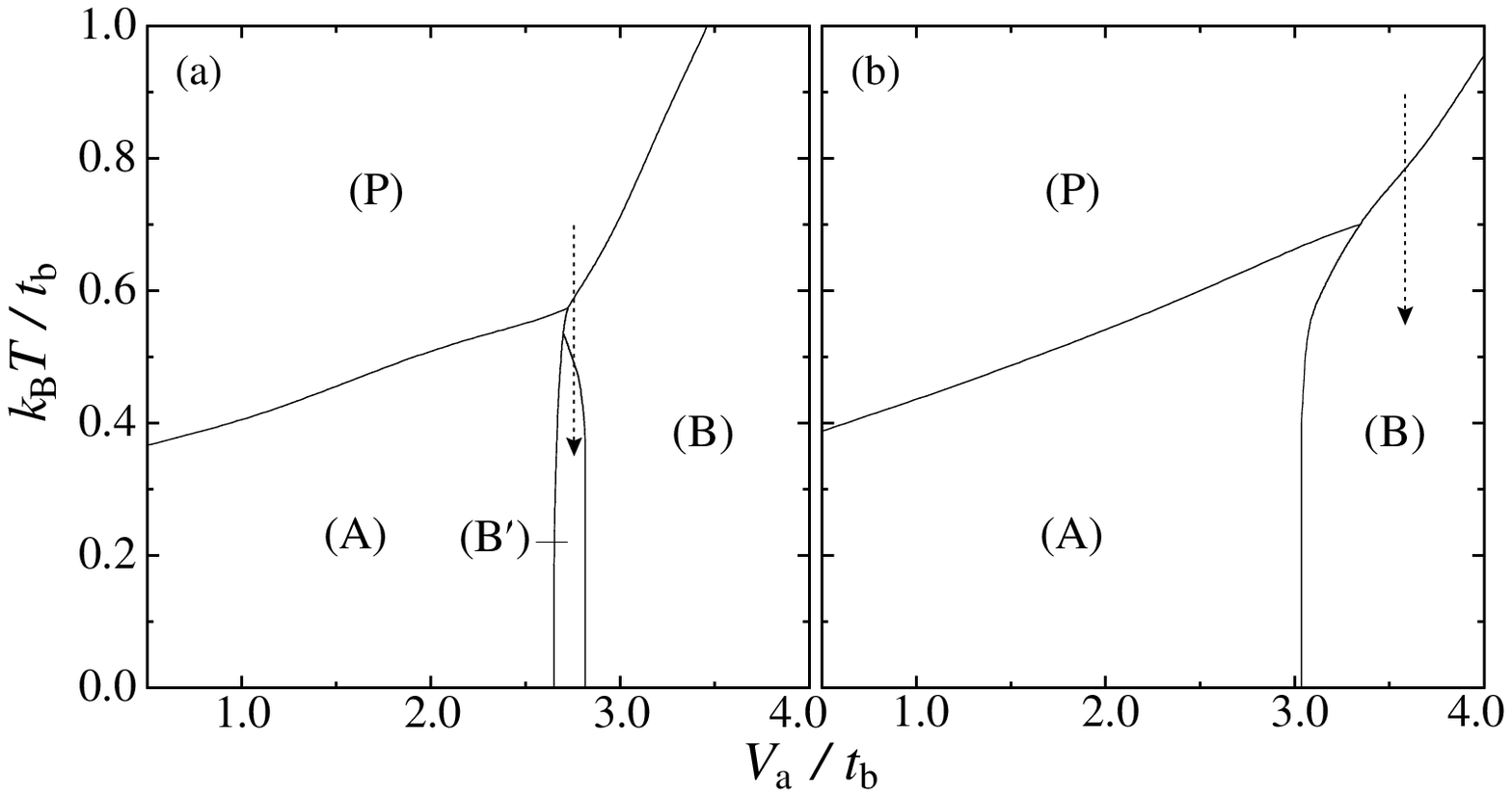,width=80mm,angle=0}}}
\vspace*{1mm}
\caption{Thermal phase diagrams on the lines of $V'/V_{\rm a}=0.75$
         in Fig. 2(a) and of $V'/V_{\rm a}=0.6$ in Fig. 2(b), where
         the dotted arrows imply the thermal behaviors observed in
         $\mbox{(TMTSF)}_2\mbox{PF}_6$ (a) and in
         $\mbox{(TMTTF)}_2\mbox{ReO}_4$ (b), respectively.}
\label{F:ThPhD}
\end{figure}
\vspace*{1mm}

  We show in Fig. \ref{F:GSPhD} ground-state phase diagrams
calculated for typical parameters \cite{K3356,K1098} of
$\mbox{(TMTSF)}_2\mbox{PF}_6$ (a) and for those \cite{P1501,N} of
$\mbox{(TMTTF)}_2\mbox{ReO}_4$ (b).
The intrinsic dimerization (bond alternation) is more remarkable in
TMTTF salts than in TMTSF salts in general.
Figure \ref{F:GSPhD}(a) may be compared with a pioneering calculation
by Kobayashi {\it et al.} \cite{K1098}, which was carried out within
the Hartree approximation under frozen lattice.
The two calculations are qualitatively alike, but the long-range
Coulomb effect of the Fock field on magnetic phases strongly reduce
the coexisting CDW and SDW phases (A$'$) and (B$'$).

   Taking account of experimental observations \cite{P1501,N}, we
focus on particular regions in Fig. \ref{F:GSPhD} and show their
thermal behaviors in Fig. \ref{F:ThPhD}.
In $\mbox{(TMTSF)}_2\mbox{PF}_6$, a precursor CDW fluctuation of type
(B) comes at about $150\,\mbox{K}$, it is reduced and a SDW
fluctuation of type (A) grows instead at about $50\,\mbox{K}$, and
then the mixed state (B$'$) is stabilized below $12\,\mbox{K}$.
Figure \ref{F:ThPhD}(a) well solves these observations.
Any transition to (B$'$), whether from (A) or from (B), is of second
order, whereas transitions between (A) and (B) are of first order.
Therefore, the $2k_{\rm F}$ CDW and SDW fluctuations may in principle
coexist in the intermediate temperature region.
It is true that we narrowly find such a multistep thermal behavior
within the present model, but we could ensure the possibility by
including the crucial effect of the anion potential \cite{R241102}.
In $\mbox{(TMTTF)}_2\mbox{ReO}_4$, on the other hand, the
lattice-tetramerized CDW state (B) is stabilized below about
$160\,\mbox{K}$.
There is neither precursor SDW fluctuation at intermediate
temperatures nor static spin ordering at low temperatures.
Figure \ref{F:ThPhD}(b) can qualitatively interpret all these
observations.
Applying pressure, or equivalently reducing bond alternation by
chemical tuning of materials, Chow {\it et al.} \cite{C1698} observed
crossover from spin-Peierls disordered states to $2k_{\rm F}$ SDW
states in the Bechgaard-Fabre salts.
Considering that in the $2k_{\rm F}$ CDW state (B) electrons become
more and more bound in singlet pairs on next-nearest-neighbor bonds
with decreasing temperature and therefore the state (B) is a fair
realization of the observed spin-Peierls phase, Figs.
\ref{F:ThPhD}(a) and \ref{F:ThPhD}(b) are widely consistent with
experiments, giving a hint in assigning parameters to the materials.

   We are grateful to Prof. Y. Nogami for fruitful discussion.
This work was supported by the Japanese Ministry of Education,
Science, Sports and Culture, and by the Sumitomo Foundation.

\widetext

\begin{references}

\bibitem{P1501}
   J. P. Pouget, S. Ravy,
      J. Phys. I 6 (1996) 1501.

\bibitem{M1522}
   S. Mazumdar, S. Ramasesha, R. T. Clay, D. K. Campbell,
      Phys. Rev. Lett. 82 (1999) 1522.

\bibitem{S805}
   H. Seo,
      J. Phys. Soc. Jpn. 69 (2000) 805.

\bibitem{M13400}
   S. Mazumdar, R. T. Clay, D. K. Campbell,
      Phys. Rev. B 62 (2000) 13400.

\bibitem{R16243}
   J. Riera, D. Poilblanc,
      Phys. Rev. B 62 (2000) R16243.

\bibitem{R241102}
   J. Riera, D. Poilblanc,
      Phys. Rev. B 63 (2001) 241102(R).

\bibitem{S1249}
   H. Seo, H. Fukuyama,
      J. Phys. Soc. Jpn. {\bf 66} (1997) 1249.

\bibitem{S235107}
   Y. Shibata, S. Nishimoto, Y. Ohta,
      Phys. Rev. B 64 (2001) 235107.

\bibitem{K3356}
   N. Kobayashi, M. Ogata,
      J. Phys. Soc. Jpn. 66 (1997) 3356.

\bibitem{K1098}
   N. Kobayashi, M. Ogata, K. Yonemitsu,
      J. Phys. Soc. Jpn. 67 (1998) 1098.

\bibitem{T125123}
   Y. Tomio, N. Dupuis, Y. Suzumura,
      Phys. Rev. B 64 (2001) 125123.

\bibitem{C1698}
   D. S. Chow, F. Zamborszky, B. Alavi, D. J. Tantillo, A. Baur,
   C. A. Merlic, S. E. Brown,
      Phys. Rev. Lett. {\bf 85} (2000) 1698.

\bibitem{N}
   Y. Nogami,
      private communication.

\end{references}
\end{document}